\documentclass[sn-mathphys-num]{sn-jnl}

\usepackage{graphicx}%
\usepackage{multirow}%
\usepackage{amsmath,amssymb,amsfonts}%
\usepackage{amsthm}%
\usepackage{mathrsfs}%
\usepackage[title]{appendix}%
\usepackage{xcolor}%
\usepackage{textcomp}%
\usepackage{manyfoot}%
\usepackage{booktabs}%
\usepackage[ruled,lined]{algorithm2e} 
\usepackage{listings}                
\usepackage{caption}                 
\usepackage{subcaption}

\usepackage{tabularx}
\usepackage{array}
\usepackage{booktabs}
\usepackage{amssymb}
\usepackage{pgfplots}
\usetikzlibrary{patterns}
\usepackage{hhline}

\theoremstyle{thmstyleone}%
\theoremstyle{thmstyletwo}%

\theoremstyle{thmstylethree}%

\begin{document}

\title[Article Title]{Fine-Grained Vectorized Merge Sorting on RISC-V: From Register to Cache}


\author[1]{\fnm{Jin} \sur{Zhang}}

\author[1]{\fnm{Jincheng} \sur{Zhou}}

\author[2,3]{\fnm{Xiang} \sur{Zhang}}

\author[2]{\fnm{Di} \sur{Ma}}

\author*[2,3,4]{\fnm{Chunye} \sur{Gong}}

\affil[1]{\orgdiv{School of Computer and Communication Engineering}, \orgname{Changsha University of Science and Technology}, \orgaddress{ \city{ Changsha}, \postcode{410114}, \country{China}}}

\affil[2]{\orgdiv{College of Computer}, \orgname{National University of Defense Technology}, \orgaddress{ \city{Changsha}, \postcode{410073}, \country{China}}}

\affil[3]{\orgdiv{Laboratory of Digitizing Software for Frontier Equipment}, \orgname{National University of Defense Technology}, \orgaddress{ \city{Changsha}, \postcode{410073}, \country{China}}}

\affil[4]{\orgdiv{National Supercomputer Center in Tianjin}, \orgaddress{ \city{Tianjin}, \postcode{300457}, \country{China}}}


\abstract{

\footnotetext{This work was completed at Laboratory of Digitizing Software for Frontier Equipment, National University of Defense Technology.}

Merge sort as a divide-sort-merge paradigm has been widely applied in computer science fields. As modern reduced instruction set computing architectures like the fifth generation (RISC-V) regard multiple registers as a vector register group for wide instruction parallelism, optimizing merge sort with this vectorized property is becoming increasingly common. 
In this paper, we overhaul the divide-sort-merge paradigm, from its register-level sort to the cache-aware merge, to develop a fine-grained RISC-V vectorized merge sort (RVMS). From the \emph{register-level} view, the inline vectorized transpose instruction is missed in RISC-V, so implementing it efficiently is non-trivial. Besides, the vectorized comparisons do not always work well in the merging networks. Both issues primarily stem from the expensive data shuffle instruction. To bypass it, RVMS strides to take register data as the proxy of data shuffle to accelerate the transpose operation, and meanwhile replaces vectorized comparisons with scalar cousin for more light real value swap. On the other hand, as \emph{cache-aware} merge makes larger data merge in the cache, most merge schemes have two drawbacks: the in-cache merge usually has low cache utilization, while the out-of-cache merging network remains an ineffectively symmetric structure. To this end, we propose the half-merge scheme to employ the auxiliary space of in-place merge to halve the footprint of na\"ive merge sort, and meanwhile copy one sequence to this space to avoid the former data exchange. Furthermore, an asymmetric merging network is developed to adapt to two different input sizes. Experiments on the RISC-V processor SG2042 show that four fine-grained optimization schemes including register strided transpose, hybrid merging network, half-merge strategy, and asymmetric merging network, improve performance by 4.05\%, 19.88\%, 12.23\%, and 11.04\% respectively. Importantly, the overall performance is 1.34x faster than the parallel sorting in the Boost C++ library, and 1.85x faster than std::sort.
}


\keywords{Parallel sort, Sorting network, SIMD, RISC-V}



\maketitle

\section{Introduction}\label{sec1}

\begin{figure}[htbp]
\centering
\includegraphics[width=32em]{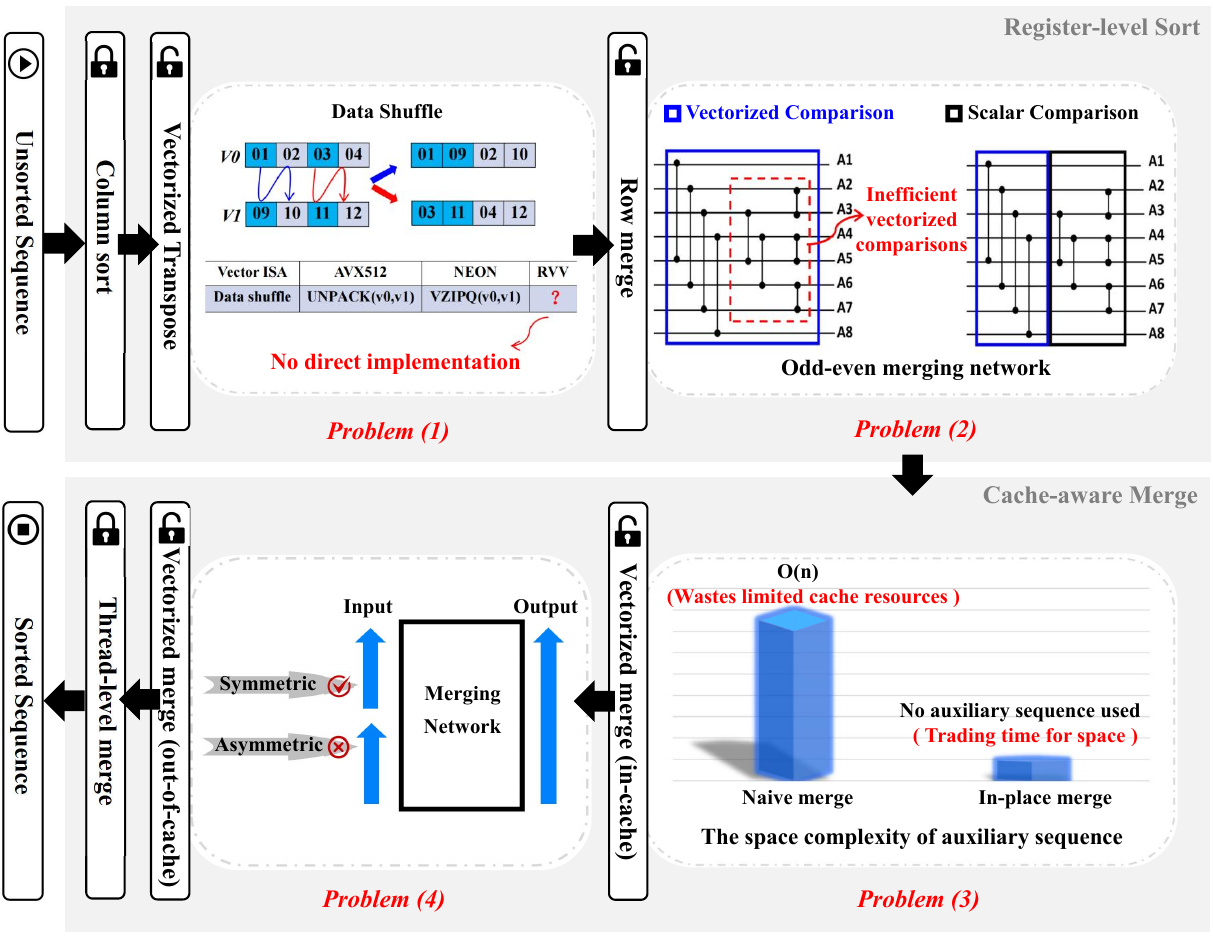}
\vspace{1em}
\caption{
The merge sort pipeline and some current existing problems: 
(1) missing the economic in-place data shuffle instruction, (2) applying expensive vectorized comparisons of the odd-even merging network for register-level sort, (3) inefficient utilization of short-supply cache resource, and (4) incompatibility between asymmetric inputs and symmetric merging network structure.
}
\label{pdrfig2}
\end{figure}

Merge sort \cite{14} is known as a divide-and-conquer algorithm. It typically decomposes a big problem recursively based on data scale into multiple small independent subproblems. Thus, most customized merge sort methods need multiple level procedures, each designed to apply modern hierarchical memory structure for high efficiency. In terms of this viewpoint, Figure \ref{pdrfig2} shows the two level pipeline of the ordinary merge sort: the \emph{register-level} sort and \emph{cache-aware} merge.

\textbf{Register-level sort} serves to sort small data whose size fits to the register width. It consists of data comparison and swapping. The comparison decides whether to carry on data swapping. So the former is pretty important. The early sort involves branch prediction and thus has to face the prediction error. In contrast, modern sort has evaded this problem because it can establish a no-branch vectorized sorting network \cite{16} via SIMD instructions. Accordingly, the sorting network by default becomes a necessity of the register-level sort. It usually has three parts: column sort, vectorized transpose, and row merge, as in the upper plane of Fig. \ref{pdrfig2}. The column sort and row merge respectively perform on different dimensions of the register group. Thus, it needs the vectorized transpose as the mediator to bridge them.

Nonetheless, it is non-trivial to implement the vectorized transpose in some instruction set architectures (ISA). Usually, two possible ways are in use. The first way to use the shuffle instruction almost retains all data transferring within the vector register, albeit reading or storing data in memory once. In contrast, the other way is relatively expensive because the gather instruction always serves to load data from the memory. The RISC-V architecture as the latest ISA is the focus of this work, but the situation above is still unchanged yet becomes more challenging because the shuffle instruction of RISC-V in itself is inefficient. As a result, this leads to inefficient implementations of the vectorized transpose ({\emph{\color{red}Problem 1}} in Fig. \ref{pdrfig2}).

Recall that vectorized sorting network serves for both column sort and row merge. The former treats each vector register as an input to the sorting network, while the latter considers each channel of the vector register as an input. The column sort is not our focus here. In the row merge, the vectorized merging network performs comparisons on the last part of the predefined sorting network. Typically, vectorized comparisons are not entirely situated in the same register channel. This has to borrow data shuffle to ensure data alignment before per comparison. As data shuffle in RISC-V is unfavorable for vectorized comparisons, it also brings into the expensive overheads when using the odd-even or bitonic merging network ({\emph{\color{red}Problem 2}} in Fig. \ref{pdrfig2}).

\textbf{Cache-aware merge} allows to merge multiple small data from the registers into larger data in the cache. It needs to handle two cases: the in-cache data and the out-of-cache data. In the first case, most vectorized in-cache merge methods \cite{30}\cite{4}\cite{8}\cite{12} introduce an auxiliary cache space to store the temporary merge results and write them back to the original cache space. This involves expensive data swapping. When the in-place merge \cite{28} works well without this auxiliary space, this hints that previous works could waste short-supply cache resources ({\emph{\color{red}Problem 3}} in Fig. \ref{pdrfig2}). However, the in-place merge has to pay for lots of data swapping. Obviously, if it is feasible to enjoy the joint strengths of both in-place merge and the auxiliary space, this could enhance cache utilization as well as balance sorting efficiency. It might be a proper solution to {\emph{\color{red}Problem 3}}.

When the to-be-merged data size comes to the cache limit, multi-way merging strategies can be used to relieve cache bandwidth bottlenecks, and remain the merge process live in the cache, thereby reducing expensive memory access. Existing works \cite{11}\cite{7}\cite{4} on multi-way merge usually combine multiple two-way merge into the binary tree-like form. In contrast, the four-way merge could be advantageous, if it can shorten the tree height, thereby reducing the merge iterations. Clearly, efficient implementation of multi-way merge is significant. The symmetric merging networks seems simple and efficient in the case merging data is just completed at the first round iteration. Unluckily, some data are out of the cache. When running this merge process for multi-way incoming data, the following iterations will become inefficient. This is because the other iterations except the first round need to receive asymmetric inputs. This incurs the incompatibility between asymmetric inputs and symmetric merging network structure ({\emph{\color{red}Problem 4}} in Fig. \ref{pdrfig2}).

In terms of the fore-said issues, a series of new insights are ready to defeat them. 
To totally remove the use of data shuffle, the transpose operation is also deemed as strided data access. An alternative is to marry strided data access with RISC-V vector extension (RVV), thereby featuring the spirit of register group. For the second issue, data shuffle is essential for vectorized comparisons, so it must be kept. According to the symmetric property of the merging network, scalar comparisons in the last rounds seem more economical because serial comparisons restrict the utilization of costly data shuffle instructions. For the in-cache merge, na\"ive merge sort wastes limited cache resources using excessive auxiliary space to store temporary merge results, while in-place merge replaces auxiliary space with plenty of data swap operations. The compromise between them could be welcome if their strengths are united with each other. The last issue is the symmetric merging structures incompatible with asymmetric inputs. The asymmetric structure is the simplest albeit easily-neglected way, but there is no direct profile of asymmetric structure for multi-way merging. Thus our solution takes a further step to extend the range of asymmetric structure for multi-way merging.

The main contributions of this paper are as follows:\par

\textbullet\, 
We explore a register-level strided transpose operation, which paves the way for efficient proxy of data shuffle on RISC-V.

\textbullet\, 
A new hybrid merging network is proposed to accelerate row merge in the register-level sort by featuring register extension as well as restrict the utilization of data shuffle instructions.

\textbullet\, 
A new merge strategy named ``half merge'' highlights the influence of auxiliary space on the customized sort by enjoying the joint strengths of both na\"ive merge sort and in-place merge.

\textbullet\, 
An asymmetric input merging network for multi-way merging is developed to increase data throughput per merge.

\section{Related Work}

This section reviews the related studies regarding the whole merge sort pipeline, which has two level procedures, i.e., register-level sort and cache-aware merge, and six subprocedures, including column sort, transpose, row merge, in-cache merge, out-of-cache merge and thread-level merge.

\subsection{Register-level sort}\label{sec2}

        \begin{figure}[htbp]
	\centering
	\includegraphics[width=26em]{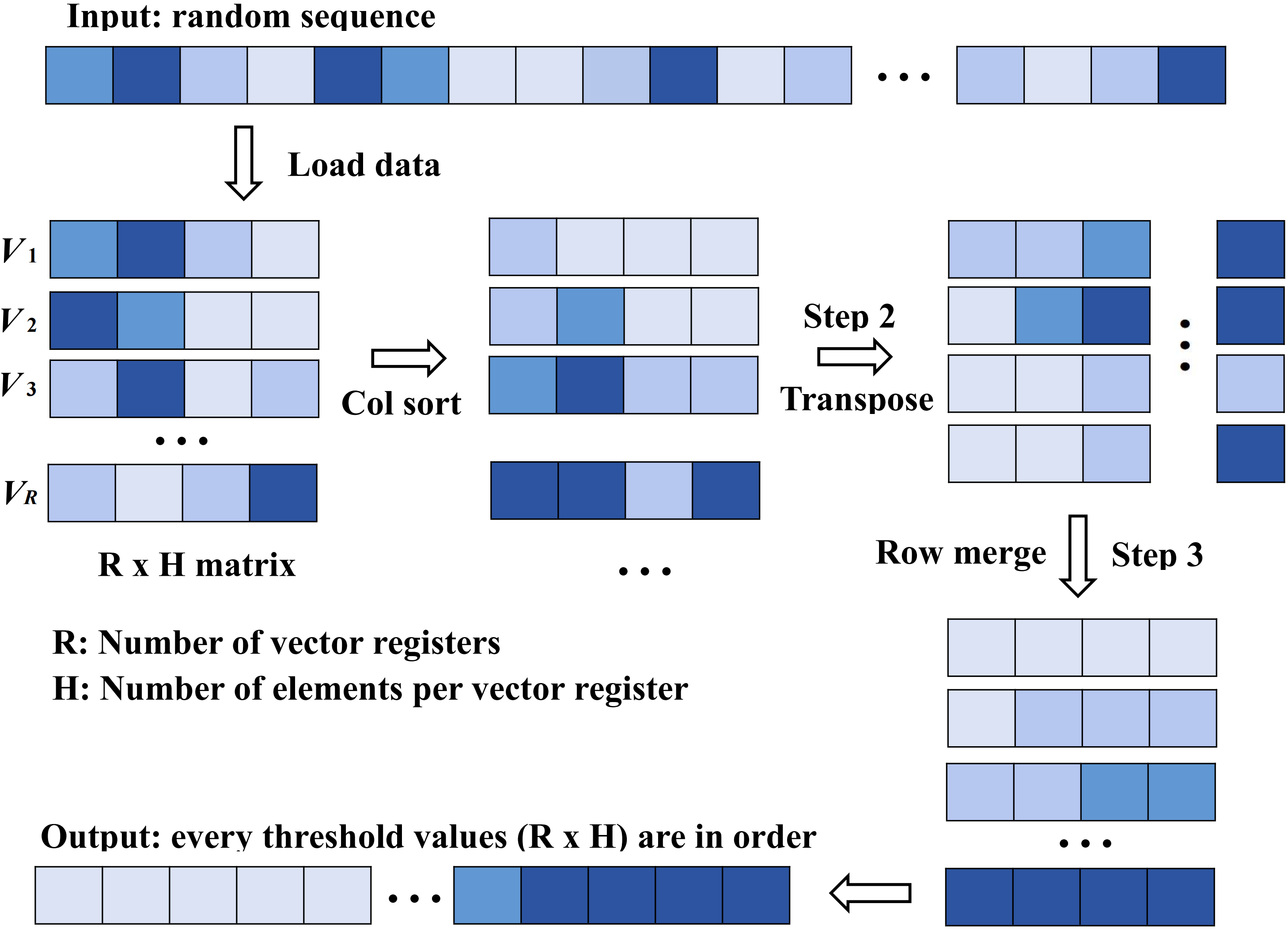}
  \vspace{1em}
	\caption{ The workflow of the register-level sort ($H$ = 4), where each square represents a data item, with darker cells indicating larger values.
}
	\label{pdrfig1}
    \end{figure}

Register-level sort by definition sorts a small-scale data within the registers. The main advantage is that it only requires reading/writing memory once, and all operations are performed on the vector register. As shown in Fig. \ref{pdrfig1}, it consists of three parts: column sort, vectorized transpose, and row merge.

\textbf{Column sort.} Column sort involves sorting the same channel across multiple vector registers, necessitating only vectorized comparisons between registers. Consequently, the quantity of vectorized comparators directly affects the sorting efficiency. In our prior work \cite{39} (implemented on NEON), we comprehensively considered two factors: register resource utilization and the simplicity of the column sorting network. Although RVV features the ability of register groups to run multiple vector register operations simultaneously, in reality, it also only provides 32 128-bit vector registers similar to ARM NEON. Therefore, we directly utilize the same strategy using 16 vector registers, and the asymmetric sorting network with fewer comparators \cite{25}. 

\textbf{Vectorized transpose.} Common vectorized transposes \cite{7}\cite{8}\cite{4} use multiple shuffle operations between registers. However, due to high cost of the RVV's shuffle instruction, it cannot directly implement vectorized transpose. We will later offer two possible solutions: one is to use a series of instructions to emulate the data shuffle, and the other is to explore RVV to find a more efficient transpose implementation.

\textbf{Row merge.} Through column sort and transpose, a $R\times H$ matrix is transformed into a $H\times R$ matrix ($H$ = 4, $R$ =16), with each row in order. To reduce unnecessary write back,  this requires the help of the vectorized merging network. For example, some works \cite{7}\cite{12} use bitonic merging network, while some \cite{8}\cite{30} utilize odd-even merging network. In our prior work \cite{39}, we analyzed the limitations of this vectorized merging network due to the inefficient data shuffle in vector instructions and proposed a hybrid merging network as an efficient solution. RVV analogously lacks efficient data shuffle operations between registers. Therefore, it motives us to use this idea of hybrid merge. However, this implementation is very different, because the hybrid strategy here is asymmetric and comes from a distinct viewpoint.

\subsection{Cache-aware merge}\label{sec2}

\textbf{In-cache merge.} After the register-level sort, each locally sorted sequence should merge into an overall block-sorted sequence. However, the length of the merge sequence might not fully load to the vector register. For further use of the small-scale SIMD sort, Inoue \textit{et al.} \cite{8} propose a idea of vectorized merge by multiple iterate merging network. Since then, vectorized merge became a replace method to serial merge. However, an aspect that deserves the alert is the use of auxiliary space. During the merge process, it is inevitable to use auxiliary sequences to temporarily store merge results. Na\"ive merge sort usually requires an auxiliary sequence of the same size as the original sequence. This results in an additional space requirement of $O(n)$, which is an obvious shortcoming. For some memory-limited machines, it can significantly influence sorting performance. Meanwhile, the use of excessively large auxiliary spaces wastes short-supply cache resources. Although \cite{28} propose an in-place merge method to no longer use the auxiliary space, this strategy of trading time for space is not well-utilized in runtime-focused customized sorting. This is why other customized sorting algorithms \cite{30}\cite{4}\cite{8}\cite{12} still do not mention this point. To this end, we design a simple yet efficient merge strategy to cooperate in-place merge with na\"ive merge sort.

\textbf{Out-of-cache merge.} Here, each thread includes multiple sorted cache blocks. These cache blocks need to be further merged until all data within the thread is sorted. Considering that the merge length exceeds the cache size, the 2-way merge is typically unsuitable due to limited cache bandwidth. Prior some work \cite{32}\cite{15}\cite{29} does not consider this issue. In contrast, other works use k-way merge to address the bandwidth bottleneck. Specifically, \cite{7}\cite{4} use a 2-way merge in each leaf node and set buffer space to let merge operation reside in cache, while \cite{11} use a 4-way merge in each leaf node to achieve better data throughput. In RVMS, our multi-way merge approach is similar to that of the latter. The core component of the multi-way merge is the merging network. However, \cite{11} only mentions that the multi-way merging network has complex dependency relationships, without further analysis of the other alternative networks. In contrast, we extend the asymmetric structure to multi-way merging networks for high performance.

\textbf{Thread-level merge.} When $T$ threads have completed sorting their allocated data, the $T$ locally sorted subsequences need to be merged to finish the final merge. To fully utilize all thread resources for parallel work, \cite{12}\cite{29}\cite{7}\cite{11} introduce a parallel partitioning strategy \cite{17}. The primary optimization involves balancing the workload to ensure that each thread can allocate a comparable amount of work. With the help of the partitioning strategy, a couple of merged sequences are assigned to multiple threads, with each thread independently being responsible for a portion of merged sequences. This is not our focus in this work.

\section{Method}\label{sec2}

\subsection{Overview}\label{sec2}

\begin{figure}[htbp]
	\centering
	\includegraphics[width=32em]{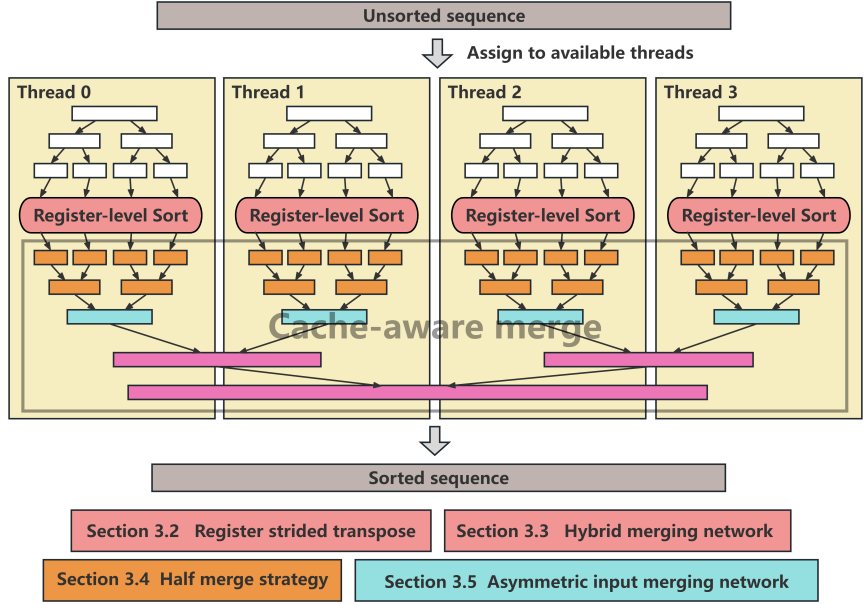}
  \vspace{1em}
	\caption{The proposed merge sort workflow with two core parts, \emph{i.e.}, the register-level sort and the cache-aware merge. The bottom part of this figure presents our improved methods for the aforementioned four questions.
}
	\label{pdrfig}
    \end{figure}

Here we propose a merge sort algorithm on RISC-V (RVMS), which follows the usual merge sort workflow. As shown in Fig. \ref{pdrfig}, it includes four stages: register-level sort, in-cache merge, out-of-cache merge, and thread-level merge. They are marked with four colors. We categorize them into two core parts: register-level sort and cache-aware merge. To fully utilize the cache feature, unsorted sequence is segmented into several blocks, with each block running in cache. These blocks are assigned among all available threads to enhance the parallelism of the merge operation. Within each cache block, the register-level sort and in-cache merge are performed to ensure that the block is sorted. Then, each block undergoes out-of-cache merge to ensure the sequences within each thread to be ordered. Finally, each thread utilizes a parallel partitioning strategy \cite{17} to collaboratively complete the merge.

\subsection{Register strided transpose}\label{sec2}

After column sort, each locally sorted sequences are distributed by column across different vector registers. This structure must be restored into its original row-wise form via the matrix transpose operation before writing the data back to memory, see Fig. \ref{pdrfig1}.
If $W < H $, the transpose of the $H \times H$ matrix can be viewed as the basic matrix transpose like the atomic operation. Thus, the transpose of an asymmetric $ R \times H$ matrix needs to first perform multiple basic matrix transposes and then adjust the positions of the vector registers.

\begin{figure}[h!]
    \centering
     \begin{subfigure}[c]{\textwidth}
        \includegraphics[width=\textwidth]{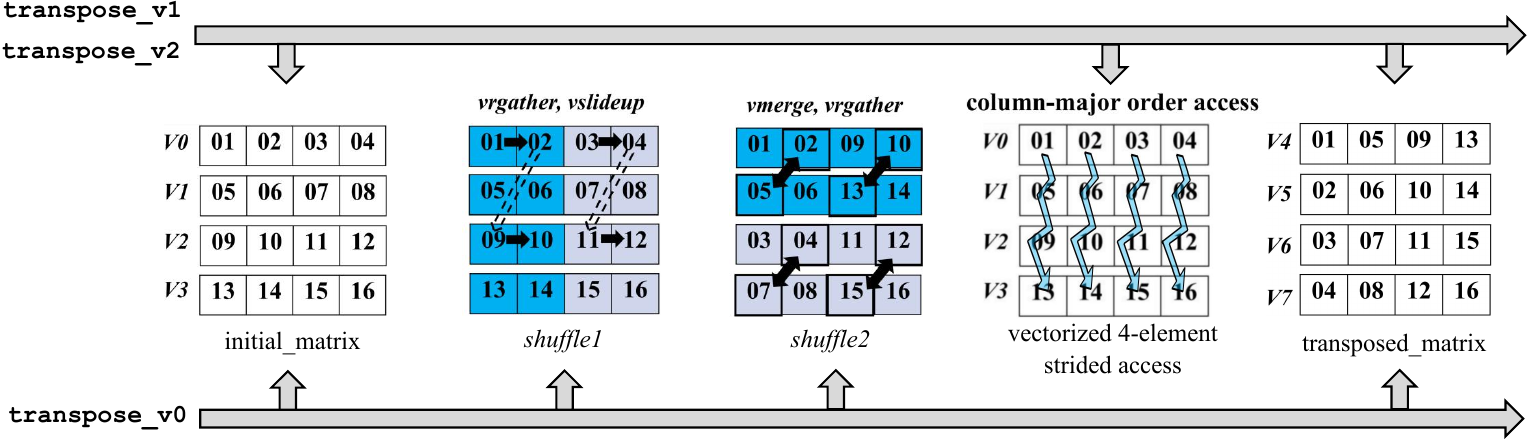}
        \captionsetup{labelformat=empty} 
        \label{fig:sub0}
    \end{subfigure}

    \lstset{
        basicstyle=\ttfamily\small,
        escapechar=@,  
        moredelim=**[is][\bfseries]{**}{**},
        xleftmargin=.12\textwidth
    }
    \begin{lstlisting}[xleftmargin=0pt]
    
// transpose_v1
    //create a register array: va
    vint32m1x4_t va = vcreate_i32m1x4(v4,v5,v6,v7); 
    //the vectorized 4-element strided access in initial_matrix
    vssseg4e32_v_i32m1x4(initial_matrix,16,va,vl);


// transpose_v2
    //create a register group: vgroup
    vgroup = vset_v_i32m1_i32m4(vundefined_i32m4(),0,v0);
    vgroup = vset_v_i32m1_i32m4(group,1,v1);
    vgroup = vset_v_i32m1_i32m4(group,2,v2);
    vgroup = vset_v_i32m1_i32m4(group,3,v3); 
    //set register group size
    size_t vl1 = vsetvl_e32m4(16);
    vint32m4_t transposed_matrix;
    //the vectorized 4-element strided access in vgroup
    transposed_matrix = vrgather_vv_i32m4(vgroup,v,vl1);


    \end{lstlisting}
    \vspace{1em}
    \caption{Three transpose implementations: \texttt{transpose\_v0} (two shuffle operations), \texttt{transpose\_v1} (memory strided operation), and \texttt{transpose\_v2} (register strided operation). The code for \texttt{transpose\_v0} is overly complex, so we will not display the complete code.
}
    \label{fig:total_image}
\end{figure}

Without loss of generality, we take $H$ = 4 for example, that is, a $4 \times 4$ basic matrix transpose will be analyzed here. As shown in Fig. \ref{fig:total_image}, the first implementation \texttt{transpose\_v0} uses other intrinsic instructions like $vmerge$, $vrgather$ and $vslideup$ to simulate the shuffle operations, the overhead is very large, compared to the shuffle instructions of the other ISAs.
By observing such two shuffle operations, they can be replaced by a common strided access operation. As shown in Fig. \ref{fig:total_image}, for a 4x4 matrix, a vectorized 4-element strided access corresponds to a column-major order access. To feature modern hierarchical memory structure, twin children of strided access are born. The first child \texttt{transpose\_v1} specifies four vector registers as a register array \texttt{vint32m1x4\_t} to store the transposed results, but such registers are mutually independent. Since the `initial\_matrix' array lives in the memory, there exists the unwelcome memory-to-register access. To avoid this point, the second kid \texttt{transpose\_v2} adopts the vector register group \texttt{vint32m4\_t} to remain the entire strided access operation active in the registers. Thus, they ensure to sort small data in registers for efficiency.

\subsection{Hybrid merging network}\label{sec2}

\begin{figure}[htbp]
\centering

\begin{subfigure}{0.49\textwidth}
  \centering
  \includegraphics[width=\textwidth]{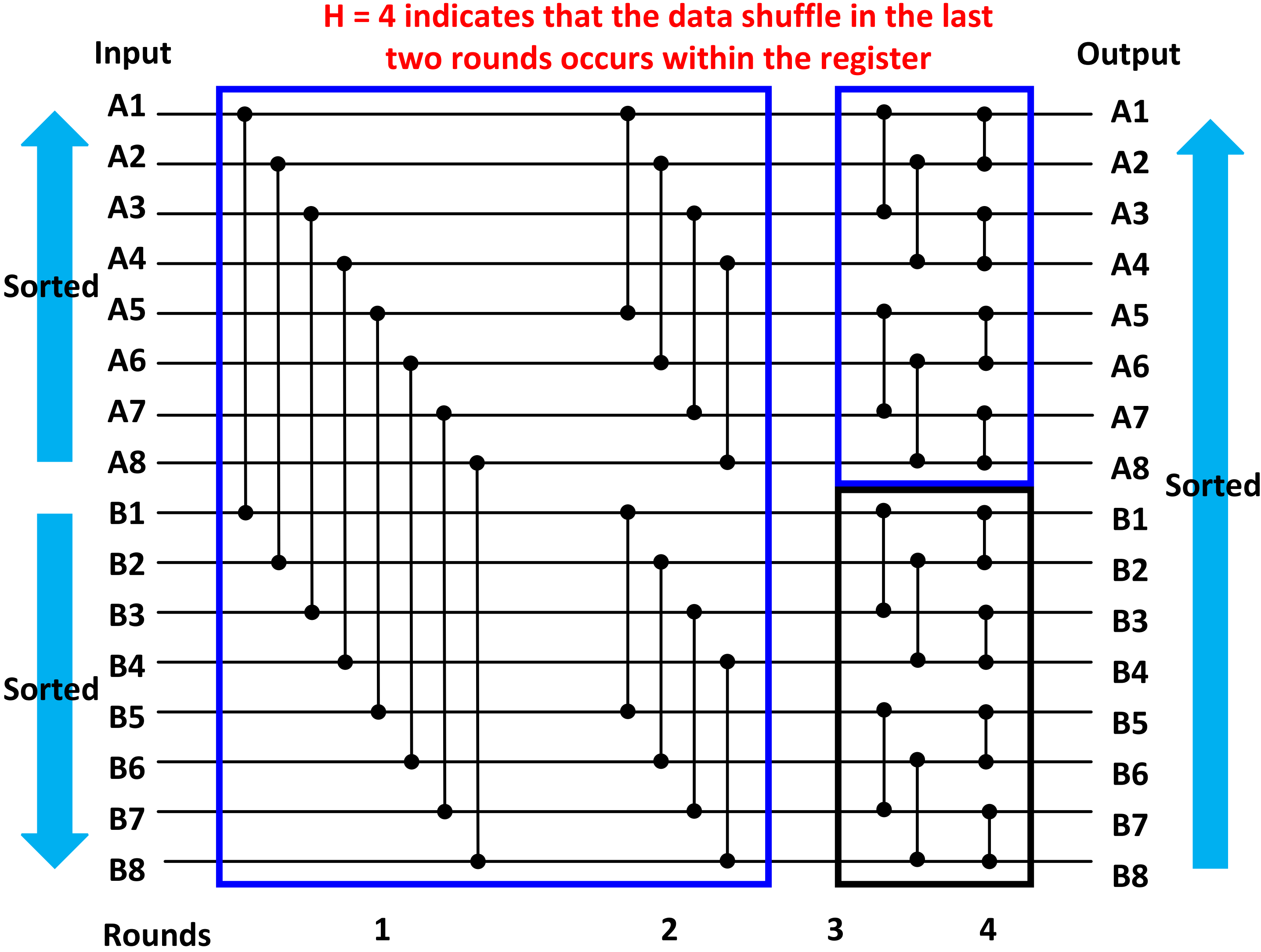}
  \caption{Bitonic}
  \label{fig:bit16}
\end{subfigure}
\hspace{1em} 
\begin{subfigure}{0.45\textwidth}
  \centering
  \includegraphics[width=\textwidth]{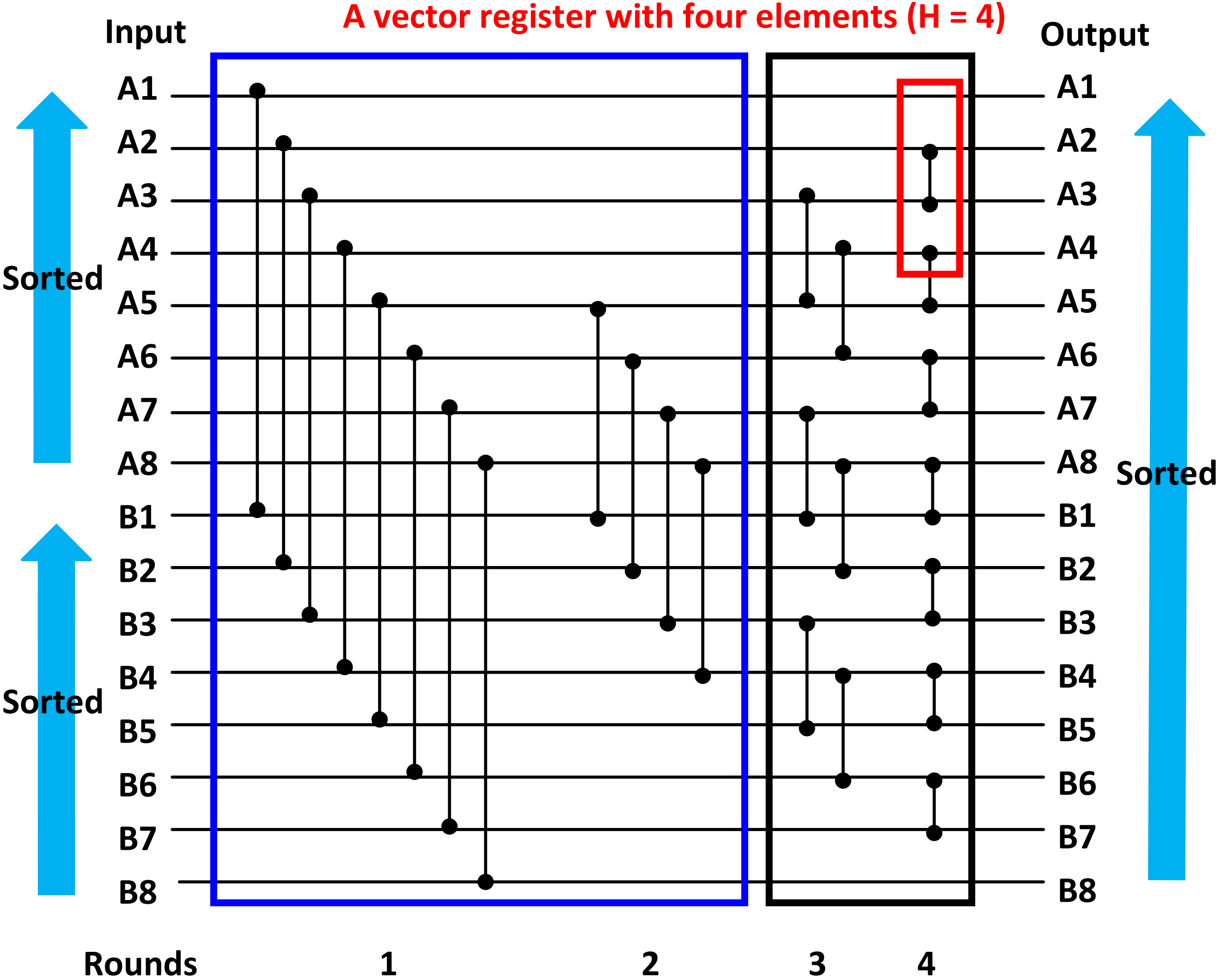}
  \caption{Odd-even}
  \label{fig:odd16}
\end{subfigure}

\begin{subfigure}{0.25\textwidth}
  \centering
  \includegraphics[width=\textwidth]{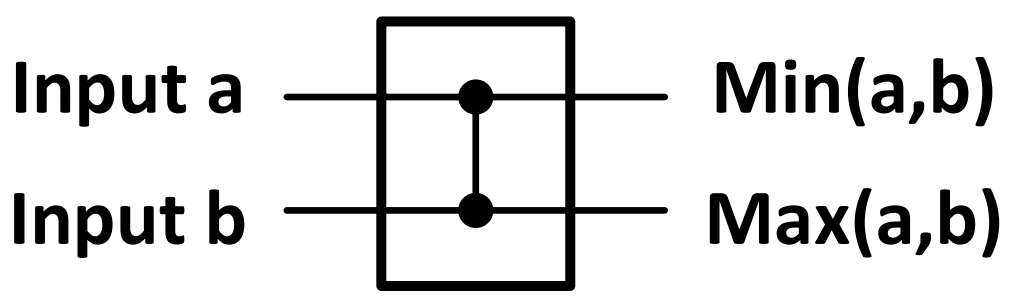}
  \caption{Comparator}
  \label{fig:4odd}
\end{subfigure}

\caption{The different hybrid strategy in 16-element bitonic and odd-even merging network, with blue and black rectangles respectively representing vectorized and serial comparisons.}
\label{fig:combined}
\end{figure}

Through the basic matrix transpose, the resultant big matrix remains to be partially in order, so a prerequisite for achieving the overall sorted $H \times R$ matrix is how to choose a favorable merging network. The usual merging network has two types: bitonic merging network, and odd-even merging network. Figure \ref{fig:combined} exemplifies their respective 16-element based structures, where a link between two black bots is a comparator (Figure \ref{fig:4odd}). Less is more---less links are more preferable. To feature vectorization in merging networks, each vectorized comparison follows a data shuffle operation for correct data alignment. In the initial several round steps, this can be accomplished by swapping the positions of vector registers. For the last several rounds, this has to apply data shuffle instructions for register-level data swapping. Actually, RVV has no direct implementation, so its implementation relies on multiple conventional instructions. Of them, the most frequently used instruction is $vmerge$, which reorders data in the registers with a mask. In a nutshell, the merging network is subject to two factors: the number of comparators, and that of the instruction $vmerge$.

According to Fig. \ref{fig:combined}, the odd-even network uses less comparators than the bitonic network. Also, from Table \ref{table:sort_comparison}, the odd-even network uses less instructions $vmerge$ than the bitonic network. Remember that the last property only holds in RVV. Some works \cite{7}\cite{12} instead utilize the bitonic network, because their vector instruction sets are not RVV, and have efficient data shuffle instructions to effectively organize more data for vectorized comparison. In terms of such analyses, the odd-even merging network is more suited here than the bitonic merging network.

\begin{table}[h!]
\centering
\caption{
   The number of $vmerge$ instructions for various merge lengths in two basic merging networks
}

\renewcommand{\arraystretch}{1.1} 
\label{table:sort_comparison}

\begin{tabular}{|c|c|c|c|}
\hline

\textbf{Merge Length}$\rightarrow$ & 4 $\rightarrow$ 8 & 8 $\rightarrow$ 16 & $2^i \rightarrow 2^{i+1}, i \geq 2$ \\ \hline
Bitonic & 6 & 12 & $3 \times 2^{i-1}$ \\ \hline
Odd-even & 2 & 4 & $2^{i-1}$ \\ \hline
\end{tabular}
\end{table}

\begin{figure}[htbp]
    \centering

    \begin{subfigure}{\textwidth}
        \centering

        \vspace{1mm}
        \begin{minipage}{.5\textwidth}
            \centering
            \begin{algorithm}[H]
                \small
                \SetKwProg{Fn}{Function}{:}{end}
                \Fn{$Comparator\_v_0$($a,l,r$)}{
                    $if$ $(a[l] > a[r])$\\
                    $\ $ $std::swap(a[l], a[r])$\;
                }
            \end{algorithm}
        \end{minipage}%
        \hfill
        \begin{minipage}{.48\textwidth}
            \lstset{basicstyle=\ttfamily\small}
            \begin{lstlisting}
             ...
11360:44b5478b lrw a5,a0,a1,2
11364:44c5470b lrw a4,a0,a2,2
11368:00f75663 bge a4,a5,11374 
             ...
            \end{lstlisting}
        \end{minipage}
        \caption{ Branch jump instruction (\textbf{bge})}
        \label{fig61}
    \end{subfigure}

    \begin{subfigure}{\textwidth}
        \centering

        \vspace{1mm}
        \begin{minipage}{.5\textwidth}
            \centering
            \begin{algorithm}[H]
                \small
                \SetKwProg{Fn}{Function}{:}{end}
                \Fn{$Comparator\_v_1$($a,l,r$)}{
                    $bool$ $flag = (a[l] > a[r])$\;
                    $int$ $temp = a[l]$\;
                    $a[l] =$ $flag?$ $a[r] : a[l]$\;
                    $a[r] =$ $flag?$ $temp : a[r]$\;
                }
            \end{algorithm}
        \end{minipage}%
        \hfill
        \begin{minipage}{.48\textwidth}
            \lstset{basicstyle=\ttfamily\small}
            \begin{lstlisting}
             ...
11338:00f726b3 slt   a3,a4,a5
1133c:42d7178b mvnez a5,a4,a3
11348:00e7a6b3 slt   a3,a5,a4
1134c:42d7178b mvnez a5,a4,a3
             ...
            \end{lstlisting}
        \end{minipage}
    \caption{ Conditional swap instruction (\textbf{mvnez})}
    \label{fig62}
    \end{subfigure}

    \begin{subfigure}{\textwidth}
        \centering

        \vspace{1mm}
        \begin{minipage}{.5\textwidth}
            \centering
            \begin{algorithm}[H]
                \small
                \SetKwProg{Fn}{Function}{:}{end}
                \Fn{$Comparator\_v_2$($a,l,r$)}{
                 \_asm\_(\\   
                       $``c.mv$ $a2,\%1 " $ \\
                        $``slt$ $a1,\%1,\%0 " $ \\
                     $``mvnez$ $\%1,\%0,a1 " $ \\
                       $``mvnez$ $\%0,a2,a1 " $ \\
                     $:``=r"(a[l]),``=r"(a[r])$ \\
                       $:``0"(a[l]),``1"(a[r])$  \\
                      $:``cc",``a1",``a2"$  \\
                     )\;
                    }
            \end{algorithm}
        \end{minipage}%
        \hfill
        \begin{minipage}{.48\textwidth}
            \lstset{basicstyle=\ttfamily\small}
            \begin{lstlisting}
             ...
113ec:863e     mv    a2,a5
113ee:00e7a5b3 slt   a1,a5,a4
113f2:42b7178b mvnez a5,a4,a1
113f6:42b6170b mvnez a4,a2,a1
113fa:4505570b srw   a4,a0,a6,2
113fe:44d5578b srw   a5,a0,a3,2
             ...
            \end{lstlisting}
        \end{minipage}
        \caption{Rewriting assembly code (\textbf{reduce a conditinal instruciton slt})}
        \label{fig63}
    \end{subfigure}
    \caption{Three different implementations of the comparator, with the right side displaying the core assembly code for the left.(\textbf{bge}$:$ \textbf{b}ranch if \textbf{g}reater than or \textbf{e}qual, \textbf{mvnez}$:$ \textbf{m}o\textbf{v}e if \textbf{n}ot \textbf{e}qual to \textbf{z}ero, \textbf{slt}$:$ \textbf{s}et \textbf{l}ess \textbf{t}han)}
    \label{111}
\end{figure}

The story does not end. Although the choice of the odd-even network is meant to use as fewer data shuffle operations as possible, this does not improve data shuffle in itself. To address this issue, our prior work \cite{39} splits the comparisons into two symmetric halves of both vectorized and serial implementations, making the instructions fully interleaved in the pipeline. This hybrid implementation drives us to further apply it for the odd-even network, but the brute force way to combine them is infeasible. This is because data comparisons within registers in the odd-even network are asymmetrical, as highlighted by the red rectangle of Fig. \ref{fig:odd16}. Thus, the way out is fully using serial comparisons in the last rounds. In serial comparisons, comparing two values needs to swap them, but this involves the conditional branch jump instruction (see Fig. \ref{fig61}). If possible, the branch misprediction should be always avoided, as it interrupts the instruction pipeline and affects the normal functioning of instructions. Fig. \ref{fig62} shows how the ternary operations replace the \texttt{if} comparisons, because a ternary operation can be forced by the compiler to become the conditional swap instruction. This implementation comes from our prior work \cite{39}. As the subfigure \ref{fig62} shows, each instruction \texttt{mvnez} follows the conditional instruction \texttt{slt}. It is clear that the \texttt{slt}s are too much redundant. If reducing such echoed conditional instructions, data swap operation will become concise and performant. This insight fits the feature of the conditional instruction in RISC-V, where the comparison result can be reused. Thus, this produces our new comparator, where an inline assembly technique serves for rewriting the comparison logic, allowing two \texttt{mvnez} instructions to share a single conditional instruction (see Fig. \ref{fig63}).

\subsection{Half merge strategy}\label{sec2}

\begin{figure}[htbp]
\centering
\begin{subfigure}{0.335\textwidth}
  \centering
  \includegraphics[width=1\textwidth]{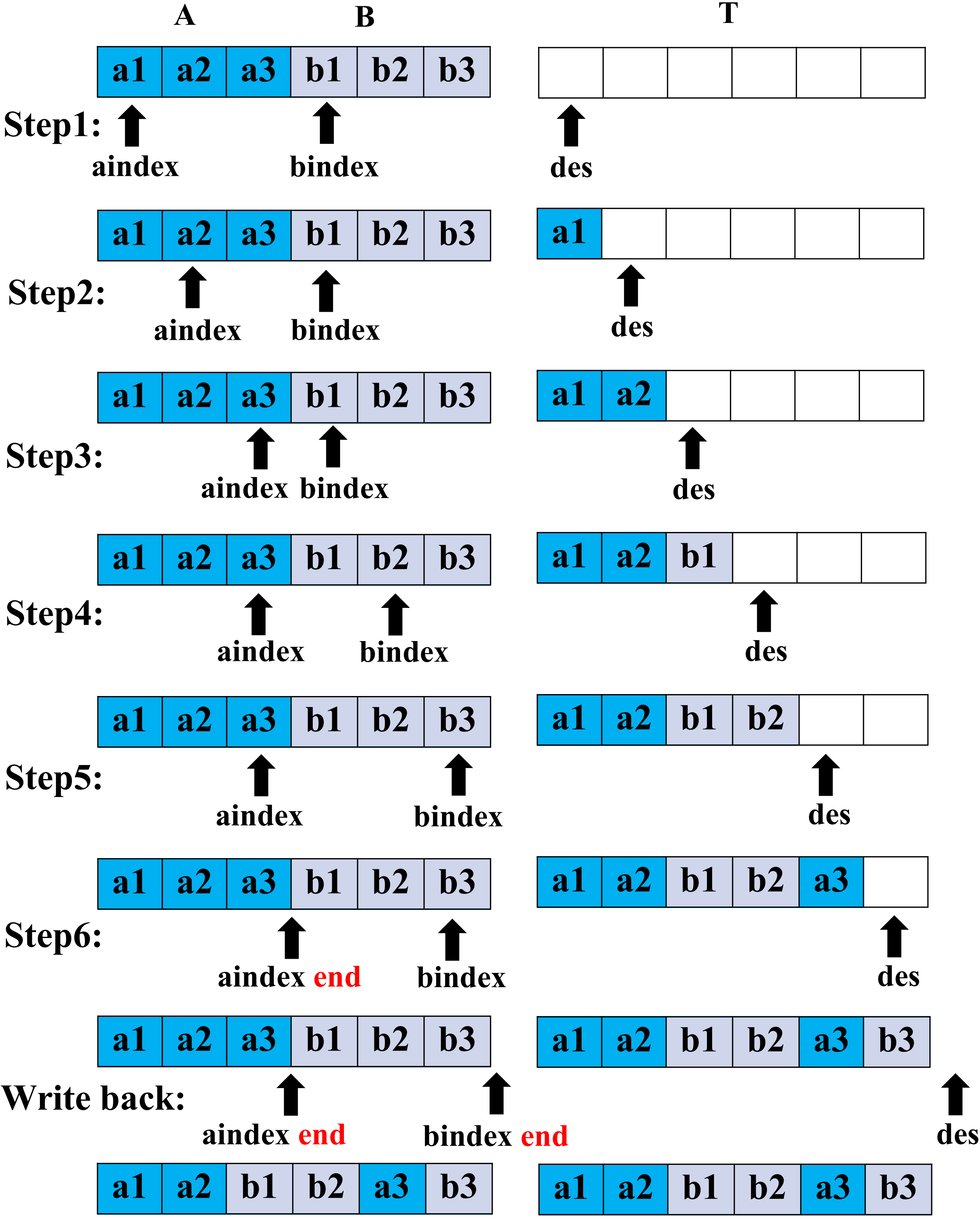}
  \caption{na\"ive merge sort}
  \label{fig:naive}
\end{subfigure}
\begin{subfigure}{0.3\textwidth}
  \centering
  \includegraphics[width=1\textwidth]{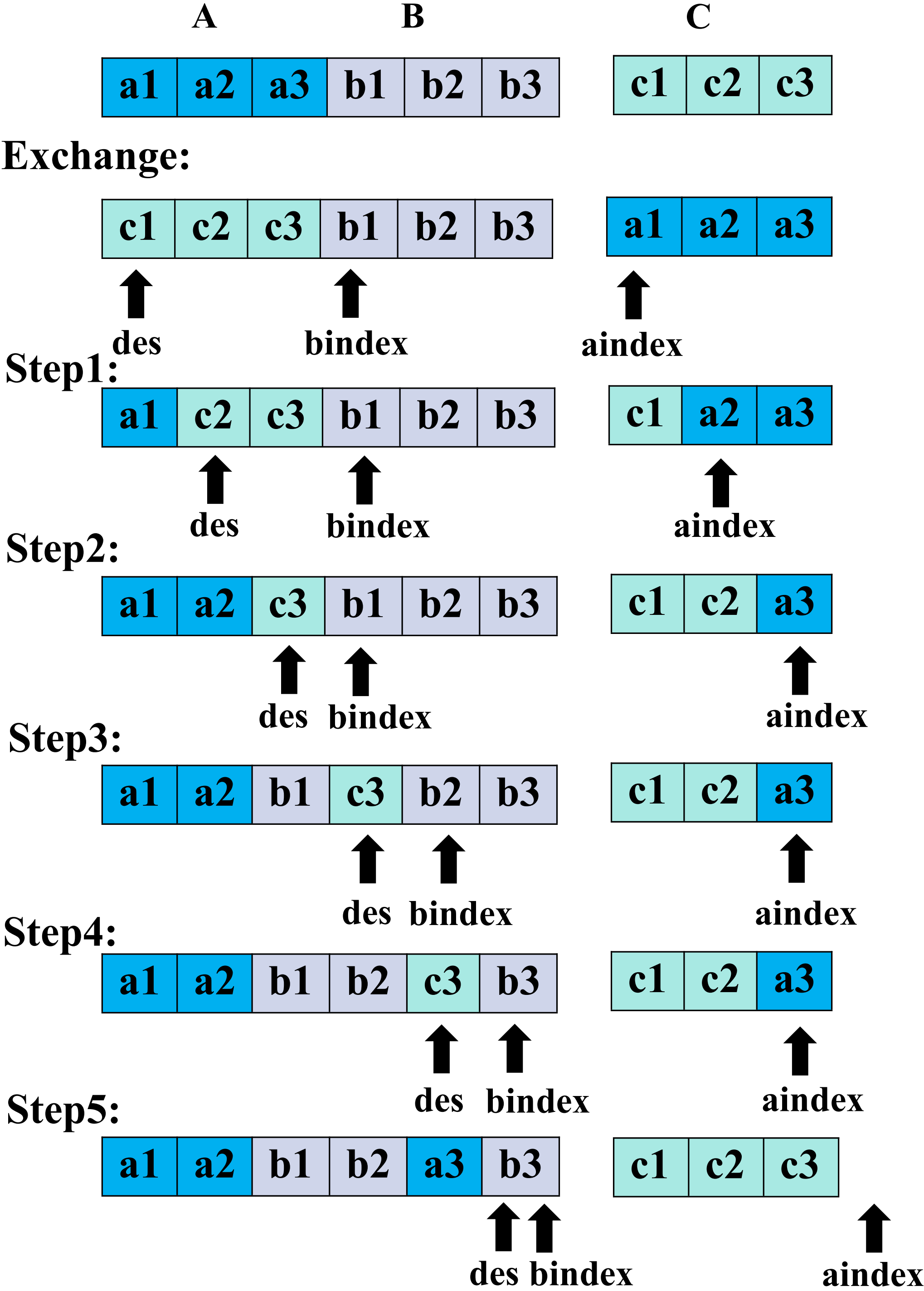}
  \caption{in-place merge}
  \label{fig:place}
\end{subfigure}
\begin{subfigure}{0.3\textwidth}
  \centering
  \includegraphics[width=1\textwidth]{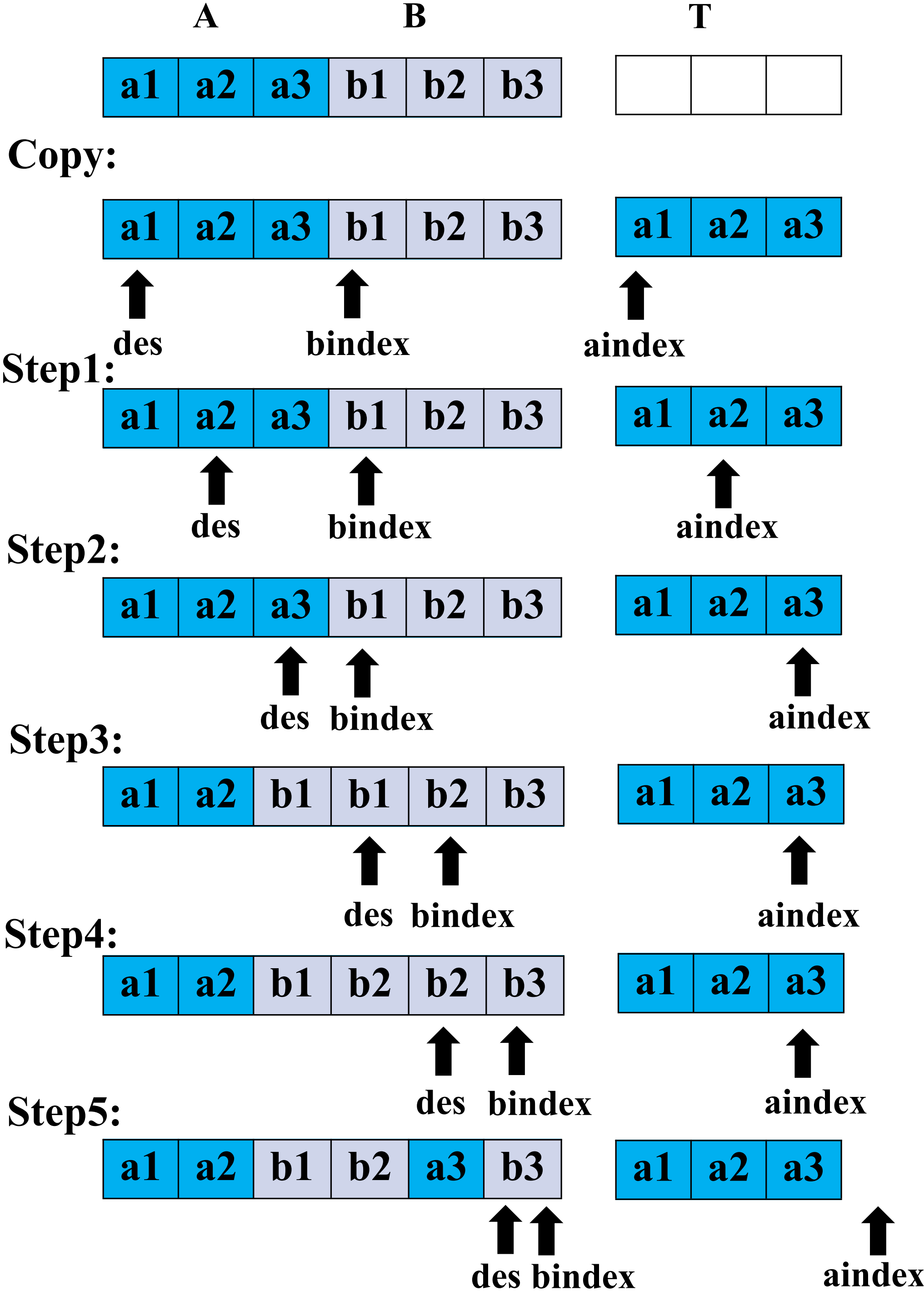}
  \caption{half merge}
  \label{fig:half}
\end{subfigure}

\caption{The flowchart of na\"ive merge sort, in-palce merge and half merge when a1 $<$ a2 $<$ b1 $<$ b2 $<$ a3 $<$ b3. }
\label{fig:combined2}
\end{figure}

Following the register-level sort, the in-cache merge will further serve to sort several small sorted data as a longer sorted sequence, but their total length has come to the register limit and possibly to the cache limit. Thus, such sorted data will be again divided into small data blocks such that their block size fits the cache width. The following merge strategy used is usually at default the na\"ive merge sort. As Fig. \ref{fig:naive} shows, when merging two data blocks $\bf A$ and $\bf B$, an auxiliary cache space engages to store their comparison outcomes. As a result, the spent cache size is the summation of the size of such two data blocks. Worsen still is consuming plenty of data comparisons and assignments. In contrast, the foresaid auxiliary space in the in-place merge is used to store another data block, whose size is half the former. As one knows, a coin has two sides. This merit pays the cost of numerous data swapping. To balance cache utilization and merging efficiency, we propose a half merge scheme to fuse the na\"ive merge sort with the in-place merge. In detail, we introduce the half auxiliary space to keep any one of two to-be-merged data blocks. Besides, during each comparison, data assignment is in use, while data swapping no longer exists. Thus the proposed half merge greatly differs from the above two merge strategies.

The above half merge strategy is not limited to serial implementation but also is suited for vectorized implementation. Given that each merge outputs the smallest $H$ elements, while the larger $H$ elements serve as input for the next iteration in a $2×H$ merging network. In serial implementation, once the auxiliary space runs out, the merge process is meant to be terminated. Instead, those larger $H$ elements need to continue to run the merging network with the other sequence. Since the in-place merge has high time cost, its vectorized implementation is computationally forbidden. In contrast, the na\"ive merge sort is tolerable in both space and time consumption. However, the na\"ive merge sort needs to use the merge results in the auxiliary space to replace the original sequence. This data transfer is expensive. Luckily, the vectorized half merge can bypass such aforementioned issues.

\subsection{Asymmetric input merging network}\label{sec2}


After the in-cache merge, each thread contains multiple sorted cache blocks. To seamlessly continue the merging process, it is necessary to carefully handle such blocks. The multi-way merge could be a feasible solution because it not only addresses the cache bandwidth bottleneck but also still remain merging operations within the cache via buffer spaces. Actually, it is non-trivial to implement it, especially for managing large data. 
To this end, we propose a multi-way merging tree, also known as the vectorized loser tree, which transforms the serial merge in each leaf node into the vectorized merging network. Each leaf in this tree is equipped with a 2k buffer space to store immediate results and performs a 4-way merge using a special 4$\times$8 merging network. 
That is, for the initial merge, the network retrieves 8 elements respectively from each of the 4 sorted sequences to perform the merge. The smallest 8 elements are the resultant sorted output, while the larger 24 elements left will continue to engage in the next merge. Consequently, the remaining structure of the merging network becomes asymmetric, whose input receives an 8-element sorted sequence and a longer 24-element sorted sequence.

\begin{figure}[htbp]
    \centering
    \begin{subfigure}[t]{0.275\textwidth}
        \centering
        \includegraphics[width=\linewidth]{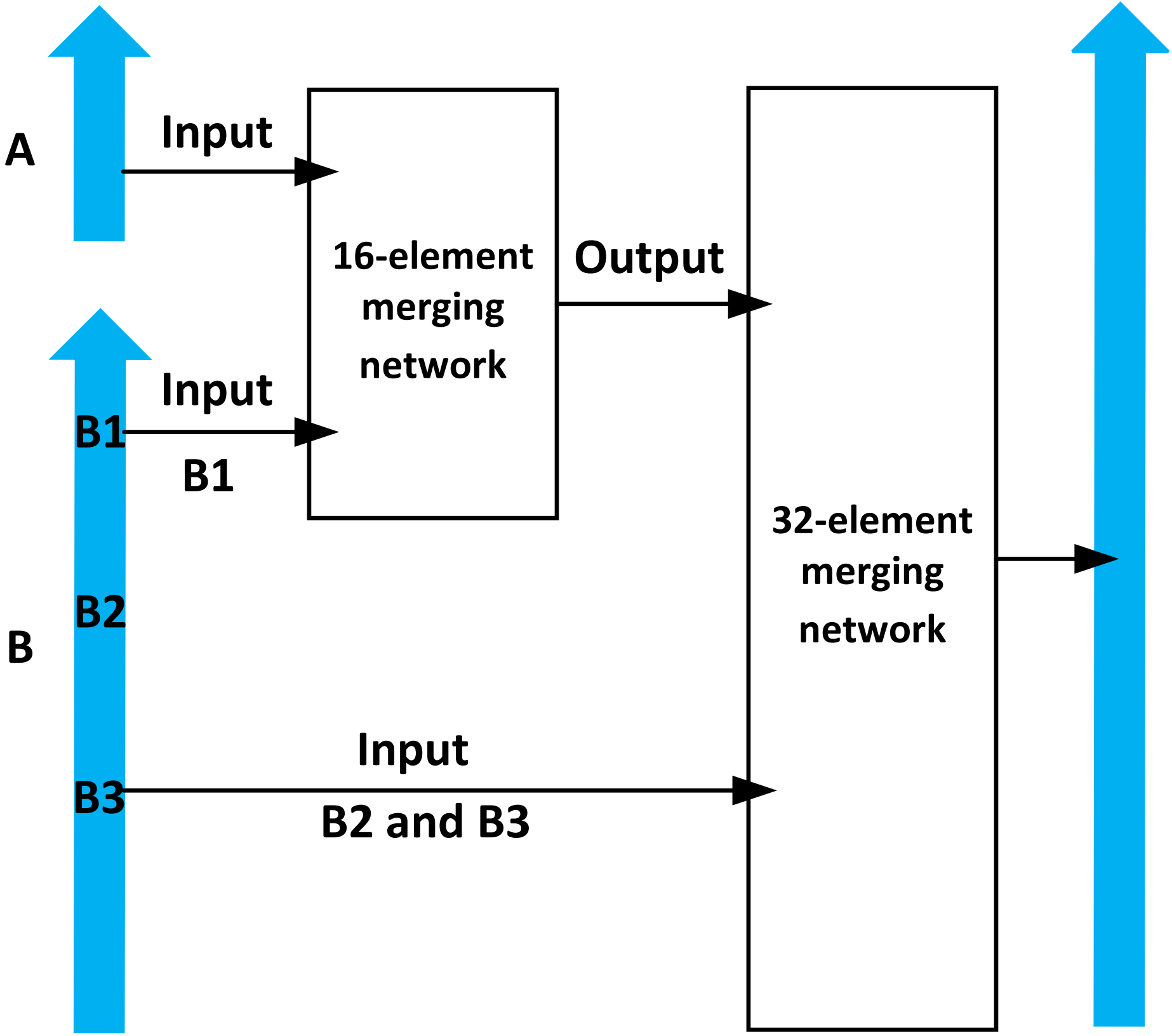}
        \caption{Standard Asymmetric}
        \label{fig:multi_merge}
    \end{subfigure}%
    \hfill 
    \begin{subfigure}[t]{0.37\textwidth}
        \centering
        \includegraphics[width=\linewidth]{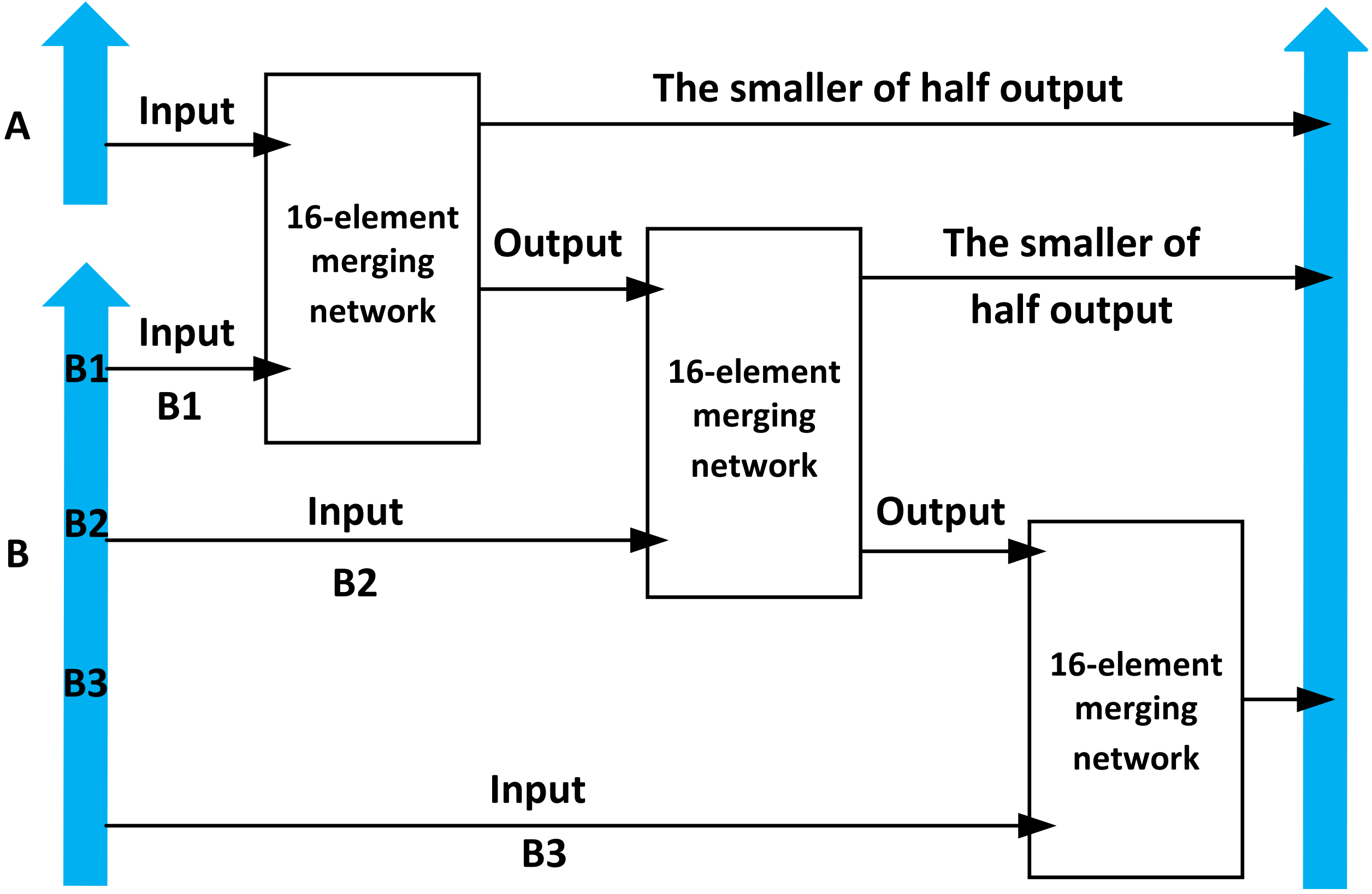}
        \caption{Iterative Asymmetric}
        \label{fig:multi_mergev1}
    \end{subfigure}%
    \hfill 
    \begin{subfigure}[t]{0.35\textwidth}
        \centering
        \includegraphics[width=\linewidth]{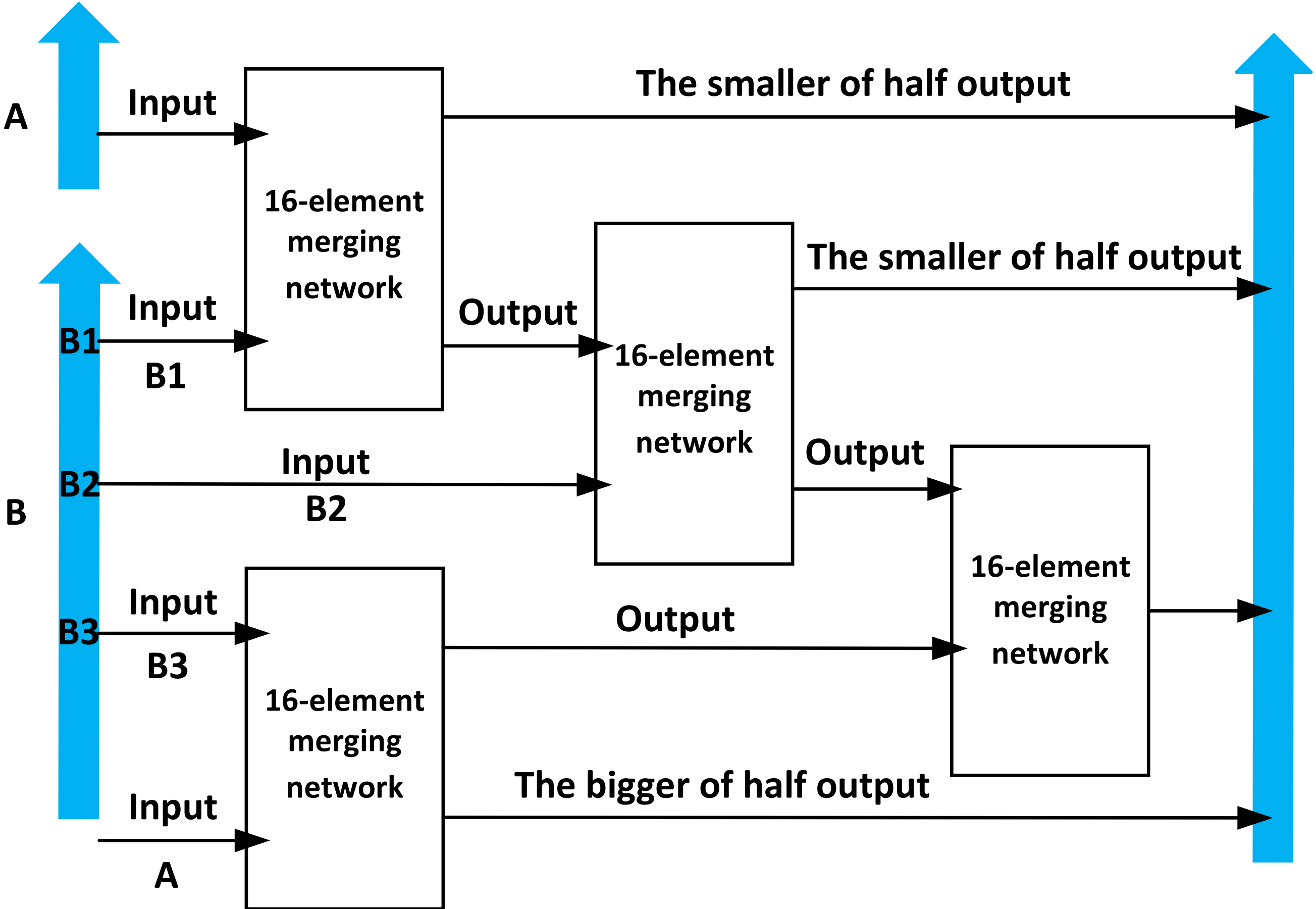}
        \caption{Parallel Asymmetric}
        \label{fig:multi_mergev2}
    \end{subfigure}
    \caption{The three different merging network structures, which take as input one 8-element sorted sequence and one 24-element sorted sequence.}
\end{figure}

Let us review this network structure. Typically, the input of a large merging network are two equal-sized smaller sorted sequences. For instance, as illustrated in Fig. \ref{fig:multi_merge}, this network serve to merge four sorted 8-element inputs. Particularly, a 16-element merging network is followed by a 32-element merging network. The former will eat two sorted 8-element sequences $\bf A$ and $\bf B1$ and then outputs a sorted 16-element sequence to the later for the final merge. Alternatively, following the same input setting, three 16-element merging networks are concatenated to achieve this goal. Although the two 16-element merging networks have fewer comparators as compared to a 32-element merging network, one network needs to wait for the outcome of the other network, as shown in Fig. \ref{fig:multi_mergev1}. To improve parallelism and reduce dependency between networks, we initially run two independent 16-element merging networks to obtain the largest and smallest 8 elements, respectively. Then the other two 16-element merging networks serve to finish the final merge in this round. In contrast with Fig. \ref{fig:multi_mergev1}, there exists an extra 16-element merging network. Since three types of asymmetric network structure have different cons and pros, how to choose the optimal network will be decided by empirical studies.

\section{Results}\label{sec2}

In this section, we evaluate RVMS on the SG2042 processor, which operates at 2 GHz, 1 MB/Cluster L2 cache, 32GB of DDR4-3200 RAM and features 64 cores. RVMS is implemented using the C language, taking advantage of RVV features for optimization. In the parallel version, we employ the OpenMP standard. All implementations are compiled using GCC 10.2.0 with \textit{-O3} level optimization. All the data used here are random 32-bit integers. Unless otherwise specified, the data scale is typically set to 100 million ($2^{27}$). Our experimental test is divided into two parts. In the localized test, we evaluate the efficacy of each improved component like transpose, comparator, merging network, merge strategy, and multi-way network structure. In the overall test, we conduct the ablation study on the overall performance of the RVMS. Subsequently, we compare the performance of the single-threaded RVMS with that of the widely used sorting function in the C++ Standard Library (std::sort) and one of the most efficient parallel sorts in the Boost C++ library (boost::block\_indirect\_sort). Finally, we assess the parallel performance of RVMS.

\subsection{Localized performance}\label{sec2}

\begin{table}[h]
\centering
\caption{The running time ($s$) of various implementation versions of transpose and comparator operations.}
\renewcommand{\arraystretch}{1.3} 
\begin{tabular}{|c|c|c|}
\hline
\textbf{Operation} & \textbf{Variant} & \textbf{Time ($s$)} \\
\hline
\multirow{3}{*}{Transpose ($trans$)} & Shuffle method ($trans\_v_0$) & 1.51 \\
& Memory strided ($trans\_v_1$) & 1.52 \\
& Register strided ($trans\_v_2$) & \textbf{0.61} \\
\hline
\multirow{3}{*}{Comparator ($comp$)} & Branch jump ($comp\_v_0$) & 2.57 \\
& Conditional swap ($comp\_v_1$) & 2.20 \\
& Assembly rewrite ($comp\_v_2$) & \textbf{1.08} \\
\hline
\end{tabular}
\label{tab:transposed_data}
\end{table}

Table \ref{tab:transposed_data} displays the optimization efficiency at each stage for both transpose and comparator operations. In the transpose operation, it is evident that $trans\_v_2$ exhibits the highest efficiency. This is because $trans\_v_2$ employs register-level stride operations, which not only circumvent the costly $vmerge$ instruction compared to $trans\_v_0$ but also reduce the times of register-to-memory accesses compared to $trans\_v_1$. For $trans\_v_1$, although it uses memory strided load to avoid the $vmerge$ instruction, it appears that the overhead of register-to-memory access is larger than that of the $vmerge$ instruction.

In the comparator operation, $comp\_v_2$ represents the final optimization version. On one hand, it eliminates the use of branch jump instructions to prevent branch prediction errors. On the other hand, it reduces extra conditional instructions by rewriting its assembly code using the inline assembly technique.
With the confirmation of the final optimization version of these base operations, we directly use their best version in the subsequent operations.

\begin{table}[h]
\centering
\caption{Merge speed (elements/$\mu s$) of different merge method in different merge size.}
\renewcommand{\arraystretch}{1.1} 
\begin{tabular}{|c|c|c|c|c|}
\hline
 & \multicolumn{4}{c|}{\textbf{Merge method}} \\ \hline 
\textbf{Size} &  \textbf{bitonic} & \textbf{hybrid bitonic} & \textbf{odd-even} & \textbf{hybrid odd-even} \\ \hline
$2 \times 4 \rightarrow 8$  & 321.56 & \textbackslash & 525.52 & \textbf{548.88} \\ \hline
$2 \times 8 \rightarrow 16$  & 76.41 & 491.31 & 635.97 & \textbf{644.09} \\ \hline
$2 \times 16 \rightarrow 32$ & 37.10 & 43.05 & \textbf{624.51} & 574.95 \\ \hline
$2 \times 32 \rightarrow 64$  & 33.40 & 36.15 & 42.06 & \textbf{449.45} \\ \hline
\end{tabular}
\label{tab:your_table_label}
\end{table}

Table \ref{tab:your_table_label} compares the merge speeds for four merge methods across different merge sizes. The results echo our previous discussions—firstly, odd-even merge is more suitable for the RISC-V architecture compared to the bitonic merge; secondly, our new hybrid strategy brings significant performance improvements. As shown in this table, the average merge speed of odd-even merge is 2.28 times faster than that of bitonic merge. This is because of the inefficiency in data swapping between vector registers in RVV, particularly with the $vmerge$ instruction. As previously mentioned, the number of $vmerge$ instructions in bitonic merge is three times greater than in odd-even merge. In the hybrid merging network, we replace vectorized comparisons with serial cousin for more light real value swap, and further optimize the serial implementation by refining the comparison logic. Interestingly, in the case of the 2x16 merge size, the merge speed of odd-even merge surpasses that of the hybrid odd-even merge. Despite our meticulous inspection of the accuracy of each local implementation in the odd-even merge process, it still presents an unexpected performance enhancement. We speculate that these enhancements may be related to the processor characteristics.

\begin{table}[h]
\centering
\caption{Percentage improvement in time of half merge compared to na\"ive merge sort.}
\renewcommand{\arraystretch}{1.2} 
\begin{tabular}{|c|c|c|c|c|c|}
\hline
\multicolumn{1}{|c|}{} & \multicolumn{5}{c|}{\textbf{Data Size}} \\ \hline
Merge method & $2^{12}$ & $2^{15}$ & $2^{18}$ & $2^{21}$ & $2^{24}$ \\ \hline
Serial merge & 36.9\% & 14.1\% & 12.2\% & 9.2\% & 10.7\% \\ \hline
Vectorized merge& 8.1\% & 3.5\% & 3.7\% & 5.3\% & 5.6\% \\ \hline
\end{tabular}
\label{tab:my_label}
\end{table}

Table \ref{tab:my_label} displays the optimization percentage achieved by applying the half merge strategy to two merging methods. As illustrated in this table, the half merge strategy exhibits varying degrees of performance improvement in different data sizes. This is because it reduces the auxiliary space usage by half, resulting in corresponding decreases in data comparison and transfer operations. In addition, it can also be observed that the percentage improvement of serial merge is greater than that of vectorized merge. This aligns with our prior discussions. In vectorized merge, each run of a $2 \times H$ merging network outputs the smaller $H$ elements, while the larger $H$ elements are used as input for the next merging network. When the elements of one sequence are used up, the larger $H$ elements should continue to run the merging network with the other sequence, instead of completing the merge as in the serial merge. So, the half merge strategy in vectorized merge only optimizes data transfer operations and does not reduce the times of data comparisons. Clearly, this difference causes the variation in percentage improvement between the two merge methods.

\begin{table}[h]
\centering
\caption{Merge speed(elements/$\mu s$) of the different merging network structure.}
\renewcommand{\arraystretch}{1.1} 
\begin{tabular}{|c|c|c|c|c|}
\hline
\multicolumn{1}{|c|}{} & \multicolumn{1}{c|}{\textbf{2-way}} & \multicolumn{3}{c|}{\textbf{4-way}}\\ \hline
\multirow{2}{*}{Network Structure} & \multirow{2}{*}{Symmetric ($v_0$)} & \multicolumn{3}{c|}{Asymmetric} \\ \hhline{~~---}
                                   &                             & Standard ($v_1$)   & Iterative ($v_2$)  & Parallel ($v_3$)  \\ \hline 
Merge Speed                        & 6.75                       & 7.32         & 7.21       & 6.64       \\ \hline
Speed-up                           & 1                          & \textbf{1.08}         & 1.06       & 0.98       \\ \hline
\end{tabular}

\label{tab:my_label2}
\end{table}

Table \ref{tab:my_label2} compares the performance of different merging network structures. As depicted in the table, $v_1$ demonstrates the best performance, being 1.08 times more efficient than the typical 2-way merge. Although the $v_2$ version involves fewer data swap operations than $v_1$, specifically comparing two 16-element networks with a 32-element network, our prior discussion (Table \ref{tab:your_table_label}) revealed that the merge speed of the latter is comparable to that of a single 16-element network from the former (odd-even). In addition, it is also observed that the performance of the $v_3$ version is slower. This indicates that adding extra operations to increase parallelism is undesirable. A detailed analysis of the results in Table \ref{tab:my_label2} reveals that the optimal merging network structure is the $v_1$ version.

\subsection{Overall performance}\label{sec2}


\begin{table}[ht]
\centering
\caption{Algorithm overall performance ablation analysis}
\label{my-label}
\setlength{\tabcolsep}{5.82pt} 
\begin{tabularx}{\textwidth}{@{}c *{9}{c}@{}} 
\toprule
Component & \multicolumn{9}{c}{Choice} \\ 
\midrule
$T0$         &   &  \checkmark  &            &                &                  &             &\checkmark&           &\checkmark     \\
$T1$          &   &              &  \checkmark &                &                  &             &\checkmark&            &\checkmark     \\
$T2$          &   &              &  \checkmark &   \checkmark  &                   &             &\checkmark&            &\checkmark     \\
$T3$          &   &              &             &                &    \checkmark     &             &           &\checkmark&\checkmark      \\
$T4$          &   &              &              &                &                  &\checkmark   &           &\checkmark&\checkmark      \\
\midrule
\textbf{Improvement(\%)} & Baseline & 4.05 & 19.88 & 18.42 & 12.23 & 11.04 & 15.54 & 18.66 & 36.06  \\
\bottomrule
\end{tabularx}
\smallskip
\par \noindent\raggedright\footnotesize  $T0:$ Register strided transpose   $T1:$ Assembly rewriting comparator   $T2:$ Hybrid network       $T3:$ Half merge   $T4:$ Asymmetric network 
\end{table}

Table \ref{my-label} demonstrates the impact of five local implementations on overall performance. As demonstrated in Table \ref{my-label}, it is evident that each local optimization contributes to significant performance improvements. Among these, the hybrid network $(T1+T2)$ exhibits the best performance enhancement. On one hand, the shuffle operation in RVV is inefficient, to this end, we propose a new hybrid merging network to accelerate by featuring register extension as well as restrict the utilization of data shuffle instructions. On the other hand, although we manually rewrite the comparator in the hybrid implementation to eliminate one extra conditional instruction per comparison, the resultant performance improvement was not substantial. Nevertheless, we still believe that assembly-level optimization represents a novel direction for custom optimization because the essence of different implementation methods for target operations is variations in instruction use. Interestingly, when coupling $T0$ with $T1$ and $T2$ together, the performance slightly falls as compared to $T1+T2$. $T0$, $T1$, and $T2$ serve for register-level sort and all work on vector registers. We speculate that $T0$ might impact the instruction pipeline of the hybrid merging network ($T1+T2$). $T0$, $T1$, and $T2$ represent optimizations for register-level sort, while $T3$ and $T4$ are related to cache-aware merge, with each set achieving significant performance improvements. Ultimately, our algorithm has achieved an overall performance increase of 36\% compared to the baseline.

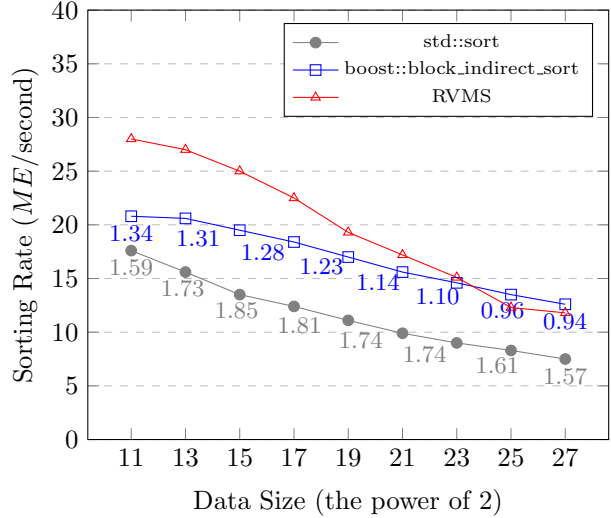
\begin{figure}[htbp] 
\centering 
\begin{tikzpicture}
\begin{axis}[
    xlabel={Data Size (the power of 2)},
    ylabel={Sorting Rate ($ME$/second)},
    ylabel near ticks,
    ylabel shift=-2pt,
    ymin=0, ymax=40, 
    ytick={0,5,10,15,20,25,30,35,40}, 
    xtick={11,13,15,17,19,21,23,25,27}, 
    legend pos=north east,
    legend style={font=\footnotesize},
    ymajorgrids=true,
    grid style=dashed,
]

\addplot[
    color=gray,
    mark=*,
] coordinates {
    (11,17.6)
    (13,15.6)
    (15,13.5)
    (17,12.4)
    (19,11.1)
    (21,9.9)
    (23,9)
    (25,8.3)
    (27,7.5)
} node[pos=0.0,below] {\scriptsize 1.59}
  node[pos=0.14,below] {\scriptsize 1.73}
  node[pos=0.29,below] {\scriptsize 1.85}
  node[pos=0.43,below] {\scriptsize 1.81}
  node[pos=0.57,below] {\scriptsize 1.74}
  node[pos=0.71,below] {\scriptsize 1.74}
  node[pos=0.86,below] {\scriptsize 1.61}
  node[pos=1,below]   {\scriptsize 1.57};

\addlegendentry{std::sort}

\addplot[
    color=blue,
    mark=square,
] coordinates {
    (11,20.8)
    (13,20.6)
    (15,19.5)
    (17,18.4)
    (19,17)
    (21,15.6)
    (23,14.6)
    (25,13.5)
    (27,12.6)
} node[pos=0.0,below] {\scriptsize 1.34}
  node[pos=0.14,below] {\scriptsize 1.31}
  node[pos=0.29,below] {\scriptsize 1.28}
  node[pos=0.43,below] {\scriptsize 1.23}
  node[pos=0.57,below] {\scriptsize 1.14}
  node[pos=0.71,below] {\scriptsize 1.10}
  node[pos=0.86,below] {\scriptsize 0.96}
  node[pos=1,below]    {\scriptsize 0.94};

\addlegendentry{boost::block\_indirect\_sort}

\addplot[
    color=red,
    mark=triangle,
] coordinates {
    (11,28)
    (13,27)
    (15,25)
    (17,22.5)
    (19,19.3)
    (21,17.2)
    (23,15.1)
    (25,12.3)
    (27,11.8)
};
\addlegendentry{RVMS}

\end{axis}
\end{tikzpicture}
\caption{Sorting Rate (ME/s: million elements per second) of different sorting methods for different data sizes. The speedup of RVMS compared to two other sorting methods is shown below the curve.}
\label{fig9}
\end{figure}

Fig. \ref{fig9} shows the performance of three sorting algorithms from 1M to 128M data sizes. This figure indicates that the overall performance of RVMS is better than other two methods. Specifically, RVMS is 1.34 times faster than block\_indrect\_sort and 1.85 times faster than std::sort at an appropriate scale. Interestingly, it is observed that the performance of RVMS gradually declines and falls below that of block\_indirect\_sort when the data scale exceeds 8M ($2^{23}$). To investigate this phenomenon further, we explore the characteristics of block\_indirect\_sort. A key advantage of this algorithm is its low memory consumption, calculated as block\_size $\times$ num\_threads. The memory consumed is directly related to the auxiliary space size mentioned earlier. The block\_size varies depending on the element size; for instance, in a 32-bit integer environment, the block\_size is set to 1024, equating the auxiliary space size to 1024 as well. Although we have proposed a half merge strategy to reduce the auxiliary space requirement by half, it remains significantly larger than block\_size $\times$ num\_threads. In large-scale data environment, this low memory consumption, accompanied by the use of a indirect pointer sorting method, substantially enhances sorting efficiency.

\pgfdeclarepatternformonly{my north east lines}{\pgfqpoint{-1pt}{-1pt}}{\pgfqpoint{10pt}{10pt}}{\pgfqpoint{9pt}{9pt}}%
{
    \pgfsetlinewidth{1pt}
    \pgfpathmoveto{\pgfqpoint{0pt}{0pt}}
    \pgfpathlineto{\pgfqpoint{9.1pt}{9.1pt}}
    \pgfusepath{stroke}
}

\pgfdeclarepatternformonly{my grid}
    {\pgfqpoint{-1pt}{-1pt}}  
    {\pgfqpoint{10pt}{10pt}}  
    {\pgfqpoint{4.5pt}{4.5pt}}
{
    \pgfsetlinewidth{0.5pt}  
    \pgfpathmoveto{\pgfqpoint{0pt}{0pt}}
    \pgfpathlineto{\pgfqpoint{4.5pt}{0pt}}  
    \pgfpathmoveto{\pgfqpoint{0pt}{0pt}}
    \pgfpathlineto{\pgfqpoint{0pt}{4.5pt}}  
    \pgfusepath{stroke}
}

\pgfdeclarepatternformonly{my backward slash}{\pgfqpoint{-1pt}{-1pt}}{\pgfqpoint{10pt}{10pt}}{\pgfqpoint{9pt}{9pt}}%
{
    \pgfsetlinewidth{1pt}
    \pgfpathmoveto{\pgfqpoint{9pt}{0pt}}
    \pgfpathlineto{\pgfqpoint{0pt}{9pt}}
    \pgfusepath{stroke}
}

\pgfdeclarepatternformonly{my horizontal lines}
    {\pgfqpoint{-1pt}{-1pt}} 
    {\pgfqpoint{10pt}{10pt}} 
    {\pgfqpoint{10pt}{3pt}} 
{
    \pgfsetlinewidth{1pt} 
    \pgfpathmoveto{\pgfqpoint{0pt}{0pt}}
    \pgfpathlineto{\pgfqpoint{10pt}{0pt}}
    \pgfusepath{stroke}
}

\pgfdeclarepatternformonly{my vertical lines}
    {\pgfqpoint{-1pt}{-1pt}} 
    {\pgfqpoint{10pt}{10pt}} 
    {\pgfqpoint{3pt}{10pt}} 
{
    \pgfsetlinewidth{1pt} 
    \pgfpathmoveto{\pgfqpoint{0pt}{0pt}}
    \pgfpathlineto{\pgfqpoint{0pt}{10pt}}
    \pgfusepath{stroke}
}

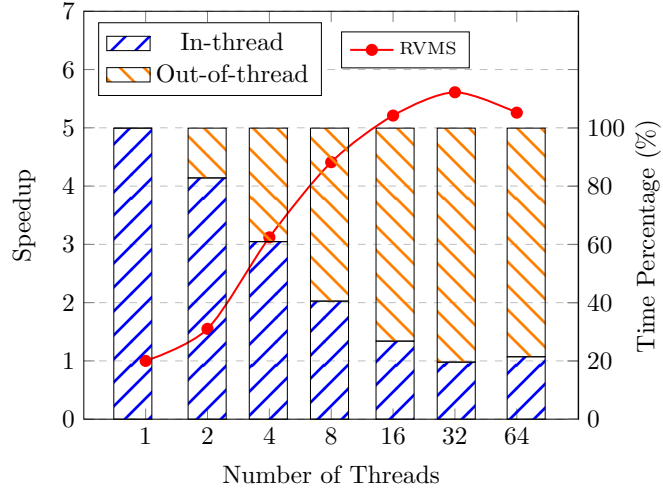
\begin{figure}[htbp] 
\centering 
\begin{tikzpicture}[scale=0.95]
\begin{axis}[
    xlabel={Number of Threads},
    ylabel={Speedup},
    ylabel style={at={(axis description cs:0.1,.5)},anchor=south},
    xtick=data,
    xticklabels={1, 2, 4, 8, 16, 32, 64},
    ymin=0, ymax=7,
    xmin=0, 
    xmax=8, 
    ytick={0, 1, 2, 3, 4, 5, 6, 7},
    legend pos=north west,
    legend style={font=\footnotesize},
    ymajorgrids=true,
    grid style=dashed,
    legend style={at={(0.65,0.86)}, anchor=south}
]

\addplot[
    color=red,
    mark=*,
    thick,
    smooth
] coordinates {
     (1,1)(2,1.55)(3,3.12)(4,4.41)(5,5.21)(6,5.61)(7,5.26)
};
\addlegendentry{RVMS}

\end{axis}

\begin{axis}[
    scale only axis,
    axis y line*=right,
    axis x line=none,
    ylabel={Time Percentage (\%)},
    ylabel near ticks,
    ylabel style={anchor=east, xshift=1.5cm,yshift=-0.2cm},
    ymin=0, ymax=100,
    ytick={0, 20,40,60,80,100},
    enlarge y limits={upper,value=0.388}, 
    width=0.5*\textwidth,
     xshift=3.1mm,
     legend style={at={(0.22,0.8)}, anchor=south}
]

\addplot[
    ybar stacked,
    bar width=15pt,
     bar shift=-5pt,
    fill=blue,
   pattern=my north east lines,
   pattern color= blue,
    fill opacity=1,
    draw=black,
    area legend
] coordinates {
    (0.5,100)(1.35,82.89)(2.05,61)(2.75,40.57)(3.5,26.82)(4.2,19.61)(5,21.45) 
};
\addlegendentry{In-thread }

\addplot[
    ybar stacked,
    bar width=15pt,
     bar shift=-5pt,
    fill=orange,
    fill opacity=1,
    pattern=my backward slash,
    pattern color= orange,
    draw=black,
    area legend
] coordinates {
      (0.5,0)(1.35,17.11)(2.05,39)(2.75,59.43)(3.5,73.18)(4.2,80.39)(5,78.55) 
};

\addlegendentry{Out-of-thread }

\end{axis}
\end{tikzpicture}
\caption{The speedup of RVMS for different numbers of threads with 128M integers. The right y-axis represents the time percentage of in-thread and out-of-thread implementations.}
\label{fig10}
\end{figure}

Fig. \ref{fig10} illustrates the parallel speedup achieved by RVMS in sorting 128M integers. The figure shows that with up to 8 threads, the speedup increases gradually, demonstrating good scalability. However, as the number of threads continues to increase beyond 8, the rate of speedup slows and may even decline. On the one hand, although the time required to calculate partition points is minimal and can often be disregarded, the synchronization overhead between threads at each parallel merge remains inevitable. On the other hand, thread-level coordination merge typically requires consideration of the tail elements of merge sequences. This merging of tail elements cannot be implemented using SIMD. As illustrated on the right y-axis of Fig. \ref{fig10}, with the increase in thread count, the proportion of time spent on out-of-thread merging grows significantly.
Although coordination merge can efficiently utilize thread resources to accelerate the merge, an excessive number of threads will lead to a decrease in the data processed per thread. Indeed, the merge speed at this scale could potentially be lower than that achieved with equal-data-scale in-thread merge. Nevertheless, multi-threaded RVMS still demonstrate good performance improvements.

\section{Conclusion}


This paper proposes a fine-grained RISC-V vectorized merge sort, named RVMS. RVMS overhauls the divide-sort-merge paradigm, from its register-level sort to the cache-aware merge. For the former, RVMS overcomes the inefficiency of the data shuffle instruction on RVV, including strides to take register data as the proxy of data shuffle to accelerate the transpose operation, and meanwhile replaces vectorized comparisons with scalar cousin for more light real value swap. For the latter, RVMS use the half-merge scheme to employ the auxiliary space of in-place merge to halve the footprint of na\"ive merge sort, and meanwhile copy one sequence to this space to avoid the data exchange. Furthermore, an asymmetric merging network is developed to adapt to two different input sizes. The results show that four fine-grained optimization schemes improve performance by 4.05\%, 19.88\%, 12.23\%, and 11.04\%, respectively. Importantly, the overall performance is 1.34x faster than the parallel sorting in the Boost C++ library, and 1.85x faster than std::sort.

\section{Acknowledgements}
This research was supported by National Key Research and Development Program of China (Grant No.2023YFB3001903) and the National Natural Science Foundation of China (Grant Nos. 62032023, 42104078 and 6190241).


\section{Declaration}

\textbf{Conflict of interest} The authors declare no competing interests.


\bibliography{sn-bibliography}

\end{document}